# Free-breathing 3D Cardiac $T_1$ Mapping with Transmit $B_1$ Correction at 3T

Paul Kyu Han[1,2]†, Thibault Marin[1,2]†, Yanis Djebra[1,2,3], Vanessa Landes[4], Yue Zhuo[1,2], Georges El Fakhri[1,2], and Chao Ma[1,2*]

[1]Gordon Center for Medical Imaging, Department of Radiology, Massachusetts General Hospital, Boston, MA, United States
[2]Department of Radiology, Harvard Medical School, Boston, MA, United States
[3]LTCI, Télécom Paris, Institut Polytechnique de Paris, France
[4]GE Healthcare, Boston, MA, United States

Running Head: 3D Cardiac $T_1$ Mapping with Transmit $B_1$ Correction at 3T

†These authors contributed equally to this work.

*Correspondence to:      Chao Ma, PhD
                         Gordon Center for Medical Imaging
                         Massachusetts General Hospital
                         125 Nashua Street, Suite 660
                         Boston, MA, 02114
                         Telephone: +1-617-643-1961
                         E-mail: cma5@mgh.harvard.edu

Word count of manuscript body: 4996

# Abstract

**Purpose:** To develop a cardiac $T_1$ mapping method for free-breathing 3D $T_1$ mapping of the whole heart at 3T with transmit $B_1$ ($B_1^+$) correction

**Methods:** A free-breathing, ECG-gated inversion recovery sequence with spoiled gradient-echo readout was developed and optimized for cardiac $T_1$ mapping at 3T. High-frame rate dynamic images were reconstructed from sparse (k,t)-space data acquired along a stack-of-stars trajectory using a subspace-based method for accelerated imaging. Joint $T_1$ and flip-angle (FA) estimation was performed in $T_1$ mapping to improve its robustness to $B_1^+$ inhomogeneity. Subject-specific timing of data acquisition was utilized in the estimation to account for natural heart-rate variations during the imaging experiment.

**Results:** Simulations showed that accuracy and precision of $T_1$ mapping can be improved with joint $T_1$ and FA estimation and optimized ECG-gated SPGR-based IR acquisition scheme. The phantom study showed good agreement between the $T_1$ maps from the proposed method and the reference method. 3D cardiac $T_1$ maps (40 slices) were obtained at a 1.9 mm in-plane and 4.5 mm through-plane spatial resolution from healthy subjects (n=6) with an average imaging time of 14.2 ± 1.6 min (heartbeat rate: 64.2±7.1 bpm), showing myocardial $T_1$ values comparable to those obtained from MOLLI. The proposed method generated $B_1^+$ maps with spatially smooth variation showing 21-32% and 11-15% variations across the septal-lateral and inferior-anterior regions of the myocardium in the left ventricle.

**Conclusion:** The proposed method allows free-breathing 3D $T_1$ mapping of the whole-heart with transmit B1 correction in a practical imaging time.

# Introduction

Cardiac $T_1$ mapping is a powerful cardiovascular magnetic resonance imaging (MRI) technique that allows quantitative assessment of tissue characteristics and underlying pathology of myocardium. Native (i.e.,without the usage of exogenous contrast agent) myocardial $T_1$ characterizes alterations in the structure and intra-/extra-cellular components of myocardium. Native myocardial $T_1$ is a well-recognized biomarker for quantitative assessment of diseases that alter tissue composition, such as iron deposition, amyloid disease, Anderson-Fabry disease, and myocarditis[1–3]. Extracellular volume fraction (ECV), measured from pre- and post-contrast $T_1$ values, provides quantitative measurement of interstitial expansion and associated diseases, such as amyloidosis, fibrosis, or myocardial edema[2,3]. ECV is an emerging biomarker for diffuse fibrosis (e.g.,in heart failure, dilated cardiomyopathy, and amyloidosis)[3], which is challenging to detect using qualitative late gadolinium enhancement (LGE) methods.

Modified Look-Locker inversion-recovery (MOLLI)[4] is a widely used method for 2D cardiac $T_1$ mapping which utilizes adiabatic inversion pulses for magnetization preparation and performs electrocardiogram (ECG)-gated balanced steady-state free precession (bSSFP) acquisitions through multiple cardiac cycles with breath-holding. Although MOLLI produces myocardial $T_1$ maps with high precision[5], the method is limited to a single-slice imaging per breath-hold. Methods have been developed to extend conventional 2D MOLLI method to multi-slice 2D or 3D acquisitions with breath-holding by leveraging the state-of-the-art parallel imaging, simultaneous multi-slice acquisition, compressed sensing, and non-Cartesian sampling techniques[6,7]. However, these methods suffer from limited through-plane resolution and coverage, often involving long or repetitive breath-holds to obtain volumetric $T_1$ maps of the heart which imposes significant burden on patients.

Various methods have been developed to overcome the limitations of breath-holding and allow 3D cardiac $T_1$ mapping with free-breathing acquisitions. Respiratory and cardiac gating-based $T_1$ mapping methods have been developed to acquire interleaved multi-slice 2D[8–10] or segmented 3D k-space data[11–13] at end-diastole with free-breathing, where effects from respiratory motion were mitigated by prospectively tracking respiratory motion using navigators or self-navigation techniques. MR fingerprinting approaches have been combined with free-breathing ECG gated acquisitions for multi-parametric cardiac MRI[14]. However, most of these methods are limited by spatial coverage, resolution in slice direction, or imaging time due to the low data acquisition efficiency of gating. Recently, free-running (i.e., no cardiac or respiratory gating) continuous acquisition methods have been proposed for 2D or 3D cardiac $T_1$ mapping[15–20]. Of note, Qi et al has reported a free-running 3D whole-heart $T_1$ mapping method[15], which uses translational respiratory motion correction and a patch-based low-rank tensor model to reconstruct 3D $T_1$ maps with isotropic resolution. The $T_1$ maps obtained by this method, however, were from 1.5T and the method may not translate well to 3T for reasons discussed below.

While 3D cardiac $T_1$ mapping methods developed up to date have been applied mostly at 1.5T, unique technical challenges arising from more severe $B_0$ and transmit $B_1(B_1^+)$ inhomogeneities need to be addressed at 3T. For instance, spoiled gradient-echo (SPGR) readout with small flip angle (FA) is often used in 3D cardiac imaging at 3T to avoid $B_0$ inhomogeneity-caused banding-artifacts associated with bSSFP readout[7,16,17,19]. However, $T_1$ mapping with SPGR readout is known to be sensitive to errors in FA caused by imperfect RF pulses and $B_1^+$ inhomogeneities[21]. The latter is particularly problematic at 3T, where $B_1^+$ variation over the left ventricle with body-coil transmission is reported on the order of 30-60%[22], leading to bias in $T_1$ estimation. Robustness of cardiac $T_1$ mapping methods with SPGR acquisitions needs to be thoroughly investigated in the presence of $B_1^+$ inhomogeneity[23,24].

In this work, we present a new cardiac $T_1$ mapping method for rapid 3D $T_1$ mapping of the heart at 3T. A free-breathing, ECG-gated inversion recovery (IR) sequence with SPRG readout was developed and optimized in terms of acquisition protocol and excitation FA for accurate and precise cardiac T1 mapping at 3T. The optimized scheme was combined with sparse $(k,t)$-space sampling along a stack-of-stars trajectory to accelerate imaging. A subspace-based image reconstruction method was used to recover high frame-rate dynamic images from highly under-sampled $(k,t)$-space data. The effects of FA errors on $T_1$ mapping were mitigated by joint estimation of $T_1$ and FA, where the reconstructed dynamic images were first binned to different respiratory motion phases and then fitted voxel-by-voxel to a signal dictionary generated using Bloch equation simulations. The effects of heart-rate variations on $T_1$ mapping were reduced by generating signal dictionary with subject-specific timing of data acquisition recorded during imaging experiment. The performance of the proposed method was characterized and validated through numerical simulations, phantom studies, and in vivo experiments on healthy human subjects (n=6). Preliminary accounts of this work have been presented previously in the form of conference abstracts(25–27).

## Methods

### Data Acquisition

The proposed ECG-gated cardiac $T_1$ mapping sequence is shown in Figure 1. A non-selective inversion pulse was applied every N+M heartbeats with two different inversion times (TI). This scheme is referred to as N-(M)-N-(M) protocol for simplicity, where N denotes number of cardiac cycles for acquisition and M denotes number of cardiac cycles for signal recovery. Data were acquired at end-diastole period using SPGR readout. A special data acquisition scheme was employed for sparse sampling $(k,t)$-space data along a stack-of-stars trajectory. A limited number of "training" data (e.g., along the $k_x$, $k_y$, and $k_z$ directions across the center of the k-space) were acquired with high sampling rate to determine the temporal changes of the underlying signal. Data at all other k-space locations were sparsely-sampled over the entire $(k,t)$-space to ensure sufficient number of measurements were acquired at each k-space location for subspace-based image reconstruction.

To track respiratory motion, 1D respiratory navigator signals were acquired in the sagittal plane at the dome of the right hemi-diaphragm at the beginning and end of data acquisition of each cardiac cycle. A spatially-selective inversion pulse was applied right after the non-selective inversion pulse to invert the magnetization signals in the same sagittal plane back to the equilibrium state and, therefore, to mitigate the contrast changes caused by the non-selective inversion pulses in the navigator signals.

### Image Reconstruction

Image reconstruction of sparsely-sampled data was performed by solving the following constrained optimization problem:

$$\hat{\rho}(x,t) = \arg\min_{\rho(x,t)} \|d(k,t) - \Omega \mathcal{F}_s\{\rho(x,t)\}\|_2^2 + \lambda \|\mathcal{F}_t\{\rho(x,t)\}\|_1,$$

$$s.t.\ \rho(x,t) = \sum_{l=1}^{L} u_l(x)v_l(t) \tag{1}$$

where $d(k,t)$ denotes the measured data, $\mathcal{F}_s$ the Fourier transform in the spatial domain, $\Omega$ the sampling mask in the ($k,t$)-space, and $\lambda$ the regularization parameter, which was chosen based on the discrepancy principle[28]. The first term of the cost function in Eq.(1) penalizes data inconsistency while second term promotes sparsity of the reconstructed dynamic images $\rho(x,t)$ in the spatio-spectral domain[29]. The constraint in Eq.(1) represents $\rho(x,t)$ as a partially-separable (PS) function[30,31], where $v_l(t)$ denotes temporal basis function, $u_l(x)$ denotes corresponding spatial coefficients, and $L$ is model order. In this work, $v_l(t)$ was estimated separately from training data using singular value decomposition (SVD) for simplified computation[29]. Image reconstruction problem was then reduced to determining the spatial coefficients $u_l(x)$ from measured data. The optimization problem in Eq.(1) was solved using an alternating direction method of multipliers (ADMM)[32] based algorithm[33]. For fast computation of $\mathcal{F}_s$, 1D Fourier transform was applied along the $k_z$ direction first and non-uniform fast Fourier transform[34] was applied in the $k_x$-$k_y$ plane for slice-by-slice reconstruction. The dynamic images were reconstructed in a coil-by-coil fashion and then combined using the sum-of-squares method to form the final reconstruction. We implemented the image reconstruction algorithm in Python and utilized the SigPy package[35] to accelerate the computation using Graphical Processing Units (GPUs). Reconstructions were performed on four NVIDIA Tesla V100 SXM2 GPUs (parallelized over slice and coil dimensions) with reconstruction time around one minute for each slice and coil.

### Estimation of T₁ and FA

Prior to parameter estimation, the reconstructed dynamic images were binned to different respiratory motion phases based on diaphragm position information extracted from navigator signals. Let $\hat{\boldsymbol{\rho}}_{n,m}$ denote the reconstructed dynamic signals at a voxel $x_n$ in a selected respiratory phase $m$ and $\boldsymbol{\eta}_{n,m} = [T_{1,m}(x_n), \theta_m(x_n)]$ denote the nonlinear parameters (i.e., T₁ and FA). We estimated $\boldsymbol{\eta}_{n,m}$ that best fit the dynamic signals using variable projection (VARPRO) method[36]:

$$\hat{\boldsymbol{\eta}}_{n,m} = \arg\min_{\boldsymbol{\eta}_{n,m}} \frac{|\hat{\rho}_{n,m}\boldsymbol{a}^{\mathrm{H}}(\boldsymbol{\eta}_{n,m})|^2}{\|\boldsymbol{a}(\boldsymbol{\eta}_{n,m})\|_2^2} \tag{2}$$

where $\boldsymbol{a}(\cdot)$ denotes an atom of a dictionary of signals calculated using Bloch equation simulation with varying $T_1$ and FA values defined on a discrete 2D grid and the actual timing of acquisition recorded during the imaging experiment. The evolution of the bulk magnetization vector was calculated excitation-by-excitation over the course of the entire scan by solving the Bloch equation numerically. The signal bases were formed using the simulated signals associated with the excitations where the spoke at the center of the *k*-space (e.g., $k_z = 0$) was acquired at each frame. The bases were then binned to a motion phase to form the dictionary for joint T1 and FA fitting. For simulation and phantom studies, dictionary was generated for a range of $T_1$ from 1-3000ms in increments of 1ms and for a range of FA from 0 to $2 \times FA_o$ in increments of $0.01 \times FA_o$, where $FA_o$ is the nominal FA, in case of joint estimation. For the in vivo study, the dictionary was generated for a range of $T_1$ from 500-2500ms in increments of 10ms and for a range of FA from $0.2 \times FA_o$ to $1.5 \times FA_o$ in increments of $0.01 \times FA_o$ in case of joint estimation, in consideration of expected smaller range of $T_1$ and FA values and for the sake of reducing computation time in dictionary generation and fitting. The time to generate the dictionary and to fit the data was 6.7±2.1min and 51.1±2.8s, respectively, using 8 Intel Xeon 2.4GHz CPUs (4-core per CPU) on a workstation.

## Simulation Study

We performed simulation studies to optimize the proposed data acquisition scheme in Fig. 1. Bloch equation simulations were performed for various ECG-gated IR schemes with following parameters in common unless otherwise mentioned: heart rate=80 bpm, acquisition window=180 ms, inversion delay times=100/180 ms (*i.e.,* delays from the inversion pulse to the beginning of the first acquisition), FA=6°, and SPGR readout. The effect of $B_1^+$ inhomogeneity on the accuracy of $T_1$ estimation was investigated for 8-8, 5-(3)-5-(3), and 10-(3)-10-(3) protocols without noise at different $B_1^+$ scenarios (i.e., $B_1^+$=0.8/1/1.2). Relative difference (in relation to the ground truth $T_1$ value) was used to assess any bias in $T_1$ estimation. The precision of $T_1$ estimation was investigated for N-(M)-N-(M) protocols with noise and perfect $B_1^+$ (i.e., $B_1^+$=1) using Monte Carlo simulations (i.e., 10,000 noise realizations). Normalized standard deviation (nSD) (i.e., standard deviation of estimated $T_1$ values normalized by standard deviation of noise and inverse of square root of acquisition time) was used to assess the precision of the $T_1$ estimation. The standard deviation of noise was assumed to be constant for all considered scenarios. Additionally, the effect of nominal FA on $T_1$ mapping was investigated for 8-8, 5-(3)-5-(3), and 10-(3)-10-(3) protocols with FAs ranging from 1 to $15°$ in $1°$ increments. Lastly, the effect of heart rate variation on $T_1$ mapping was investigated for 10-(3)-10-(3) protocol with heart rate ranging from 50 to 120bpm in 5bpm increments.

Note that the $(k,t)$-space is highly undersampled and the dynamic images are reconstructed with an explicit low-rank constraint in the proposed method. Therefore, different from MOLLI where the $k$-space of each frame can be fully sampled, the imaging time of the proposed method is given by $T_{Proposed} = \frac{P_s P_z L}{Q_c} \times t_{RR} \times \frac{M+N}{N}$, where $P_s$ and $P_z$ denotes the number of spokes in the $k_x$-$k_y$ plane and phase encoding steps in the $k_z$ axis, respectively, which are determined by the required spatial resolution; $L$ is the rank of the dynamic images; $t_{RR}$ denotes a fixed cardiac cycle duration; and $Q_c$ denotes the number of spokes acquired per cardiac cycle. Here, $P_s P_z L$ is the total number of unknowns of the low-rank model when the temporal basis is predetermined. Note that the inherent rank of the dynamic images reflects the spatial-temporal correlations of the temporal signal variations of all the voxels. It depends on the respiratory motion pattern and the $T_1$ value distributions rather than protocol parameters N and M. Since $t_{RR}$, $P_s$, $P_z$, $L$ and $Q_c$ were the same for different imaging protocols, the imaging time of the proposed method was calculated as $T_{Proposed} = \frac{M+N}{N}$ in our simulation study for simplicity.

## *Phantom Study*

A structured phantom consisting of 21-vials of deionized water doped with concentrations of gadolinium (Dotarem®) varying from 0 to 0.5mmol/L was built to validate the performance of the proposed method. Imaging experiment was performed using a 3T MR scanner (MAGNETOM Trio, Siemens Healthcare, Erlangen, Germany) with a body-coil for transmission and a 12-channel head coil for reception. Acquisitions were performed using 8-8, 5-(3)-5-(3), and 10-(3)-10-(3) protocols for a simulated heart rate of 80bpm. Common imaging parameters were: field-of-view (FOV)=360×304mm$^2$, matrix-size=192×162, slice-thickness=6mm, FA=$9°$, TR/TE=3.0/1.5ms, inversion delay times=100/180ms, and SPGR readout. Fully-sampled $(k,t)$-space data were acquired on a Cartesian grid with a temporal resolution of 30 ms (*i.e.*, 10 phase-encoding [PE] lines per frame) to evaluate the performance of different acquisition protocols. The 8-8, 5-(3)-5-(3), and 10-(3)-10-(3) protocols were repeated over 400, 400, 640 cardiac cycles, respectively, to ensure full-sampling of the (k,t)-space data. An IR sequence with fast spin echo (FSE) readout was performed to obtain reference $T_1$ maps with the following imaging parameters: FOV=360×304mm$^2$, matrix-size=192×162, slice-thickness=6mm, TR=10,000ms, echo-train-length=7, and TI=50/100/250/500/750/1000/1500/2000/2500/3000ms. An additional scan with MOLLI (37) was performed for comparison with the following parameters: FOV=360×304mm$^2$, matrix-size=192×162, slice-thickness=6mm. Region-of-interests (ROIs) were drawn within each vial and the average $T_1$ value within each vial was used for analysis. Scatter plots were generated to show the correlation between the $T_1$ values from the different methods and those from the reference IR-FSE method. Bland-Altman analysis was performed to analyze the agreement between the $T_1$ values from the different methods and those from the reference IR-FSE method.

## In Vivo Study

Six healthy volunteers (four males and two females; 32 ± 3 years) were recruited under a study protocol approved by our local Institutional Review Board (IRB). Written informed consent was obtained from all subjects before study participation. Imaging experiments were performed using a 3T MR scanner (MAGNETOM Trio, Siemens Healthcare, Erlangen, Germany) with a body-coil for transmission and spine and surface coils for reception. Imaging was performed using 10-(3)-10-(3) protocol with data sampling following a stack-of-stars trajectory. Two frames were acquired per cardiac cycle, each consisting of k-space spokes along the same angle in the $k_x$-$k_y$ plane over all $k_z$ encodings and 3 additional training lines at the center of the k-space along the $k_x$, $k_y$, and $k_z$ direction, respectively (Fig.1). The spoke angle varied randomly from frame to frame following uniform random distribution. The other imaging parameters were: FOV=308×308×180mm$^3$, matrix-size=160×160×40, image orientation=short-axis view, FA=9°, TR/TE=3.4/1.7ms, and inversion delay times=100/180ms. A relatively large through-slice coverage was chosen to mitigate errors in FA due to imperfect slab excitation profile in the presence of both respiratory and cardiac motions. Data acquired over the first 800 cardiac cycles (corresponding to anticipated scan time of 10-min considering average adult heart rate of 80 bpm) were used for reconstruction and analysis. The dynamic images were reconstructed using temporal basis functions $v_l(t)$ estimated from training data with model order L=15. The model order was chosen based on the singular value decay of the Casorati matrix formed by the training data as in the previous work on using low-rank constraints for image reconstruction(29,33). The reconstructed dynamic images were then binned into 8 respiratory motion bins using the acquired 1D respiratory navigator signals. The number of respiratory bins was chosen to be 8 based on our previous experience on using MR for respiratory motion correction in PET(33). The diaphragm position was first estimated for each frame by fitting a logistic function to the 1D spatial profile near the interface between the liver and lung. Frames were then grouped into bins according to the estimated diaphragm position while ensuring a similar number of frames within each bin. Results obtained from respiratory motion phase at or near end-exhalation were used for analysis. The short-axis view slices were divided into ROIs of 16 segments according to the AHA recommendations(38) for analysis. For comparison, 2D $T_1$ maps were acquired using MOLLI(37) for five slices in the short-axis view over the apical, mid-cavity, and basal regions of the heart and for one slice in the long-axis 4-chamber view with the following parameters: FOV=360×304mm$^2$, matrix-size=192×162, slice-thickness=4.5mm. All the five short-axis slices were categorized into apical, mid-cavity, and the basal regions based on location and were used for analysis. The mean and standard deviation of the $T_1$ and $B_1^+$ (defined as the ratio between the measured and nominal FAs) were calculated for each ROI and were visualized through bull's-eye plot and bar plot. Statistical analysis was performed using Wilcoxon signed rank test to compare $T_1$ values obtained by MOLLI and the proposed method.

## Results

Results from the simulation study are shown in Figures 2 and 3. When estimating $T_1$ only with the assumption of perfect $B_1^+$ (i.e., $B_1^+$=1), noticeable bias in $T_1$ estimation was found in all the investigated imaging protocols in the presence of typical $B_1^+$ inhomogeneities at 3T (the blue dashed lines in Fig.2). The 8-8 protocol, which has the highest data acquisition efficiency (i.e., number of k-space lines acquired per unit time) among all the schemes, showed the highest sensitivity to $B_1^+$ inhomogeneity (Fig.2a). Insertion of cardiac cycles for signal recovery reduced this bias in $T_1$ estimation at the cost of imaging time (Figs.2b and c). Joint $T_1$ and FA estimation led to unbiased $T_1$ estimation in the simulation study as expected (the red solid lines in Fig.2).

Figure 3a shows the precision of $T_1$ estimation for different imaging protocols. The nSD of the proposed method with joint $T_1$ and FA estimation was minimized and plateaued around N=13, 14, 14, and 13 for N-N, N-(1)-N-(1), N-(2)-N-(2), and N-(3)-N-(3) protocols, respectively. The nSD of the proposed method with $T_1$ estimation only was minimized and plateaued around N=12, 8, 10, and 12 for N-N, N-(1)-N-(1), N-(2)-N-(2), and N-(3)-N-(3) protocols, respectively. While resulting in unbiased estimation of $T_1$, joint $T_1$ and FA estimation led to larger nSD than the case of estimating $T_1$ only. Increasing the number of cardiac cycles for signal recovery reduced the nSD observed in joint $T_1$ and FA estimation, eventually to a level similar to those observed in the case of estimating $T_1$ only with 10-(3)-10-(3) protocol. Note that joint $T_1$ and FA estimation was unstable for 8-8 protocol despite the desired data acquisition efficiency. Figure 3b shows effects of FAs on the precision of $T_1$ estimation for different imaging protocols. The nSD was minimized and plateaued around FA of 9° in the case of joint $T_1$ and FA estimation for both 5-(3)-5-(3) and 10-(3)-10-(3) protocols. Based on the results, FA of 9° was used for the ECG-gated IR sequence with SPGR readout in the phantom and in vivo experiments. Figure 3c further shows that heartbeat rate has only marginal effects on precision of $T_1$ estimation for 10-(3)-10-(3) protocol, which is the protocol selected for in vivo experiments.

The results from phantom studies are shown in Figure 4 and Supporting Information Figures S1 and S2. Banding artifacts were observed in the estimated $T_1$ maps from MOLLI, whereas no noticeable artifacts were shown in the estimated $T_1$ maps from proposed method (Fig.4a). 5-(3)-5-(3) and 10-(3)-10-(3) protocols both showed accurate $T_1$ mapping in relation to the reference IR-FSE method when $T_1$ and FA were estimated jointly, as shown in the correlation plots in Fig.4b and Bland-Altman plots in Fig.4c. 8-8 protocol produced $T_1$ maps with large variations when $T_1$ and FA were estimated jointly (Fig.4a), which matched with the simulation results in Fig.3. Supporting Information Figure S1 shows the results obtained from the same experiment but with estimation of $T_1$ only assuming prefect $B_1^+$ (i.e., $B_1^+ = 1$). $B_1^+$ inhomogeneity caused bias was found in the estimated $T_1$ maps from IR sequence with SPGR readout as expected. Compared to the case of estimation of $T_1$ only, joint $T_1$ and FA estimation reduced the limits of agreement by 77.7, 44.7, and 49.4 ms for the $T_1$ mapping experiment with 8-8, 5-(3)-5-(3), and 10-(3)-10-(3) protocol, respectively (Figs.4c and S1c). Supporting Information S2 shows results from another phantom study with variation in $B_1^+$ field strength via control of transmitter voltage. 5-(3)-5-(3) protocol achieved robust $T_1$ mapping when $T_1$ and FA were estimated jointly, despite the variation in $B_1^+$ field strength. Estimated $T_1$ from each vial were in good agreement with those estimated from reference IR-FSE method. Ratios between different nominal FAs and estimated average FAs from each vial were also in good agreement.

Results from the in vivo study are shown in Figures 5 to 10 and Supporting Information Figures S3 and S4. The average heartrate of the 6 volunteers was 64.2±7.1bpm (min:53.7bpm, max:73.5bpm). The acquisition time for the 6 volunteers was 14.2±1.6min (min:12.2min, max:16.4min). Figure 5 shows representative reconstructed images at various slice positions and inversion times from Subject 1 using the proposed method. No significant artifacts were seen in the reconstructed images across different slices and TI times. After respiratory motion binning, an average of 153.8±23.4 inversion times were observed per respiratory motion bin over the course of 800 heartbeats across all subjects. Figure 6 shows representative short-axis-view $T_1$ maps from Subject 1 and 2 obtained by the proposed method. Figure 6 also shows 4-chamber-view $T_1$ maps which were generated by re-slicing the 3D $T_1$ map from the proposed method. For comparison, $T_1$ maps from MOLLI at the same slice position and orientation are shown at the bottom of each subfigure. Overall, the $T_1$ maps from the proposed method were comparable to those from MOLLI. Note that the nominal in- and through-plane resolution from the proposed method was 1.9mm and 4.5mm, respectively. Figure 7 further shows 3D $T_1$ maps from Subject 2, covering the whole left ventricle from base to apex. As can be seen, high-quality 3D $T_1$ maps of the heart were produced using the proposed method. Figure 8 shows a quantitative comparison of $T_1$ maps from all subjects using the proposed method (with and without joint $T_1$ and FA estimation) and MOLLI, respectively. The bull's-eye plot and bar plot of the mean and standard deviation of $T_1$ values from each ROI show a very good agreement between the two methods. This observation is also supported by the Bland-Altman plots shown in Supporting Information Figure S3. Compared to the case of estimation of $T_1$ only, joint $T_1$ and FA estimation reduced the standard deviation of $T_1$ values across 16 myocardial segments by 30.5ms . Statistical test showed that the $T_1$ values of the 16 myocardial segments from the proposed method with joint $T_1$ and FA estimation were not statistically different from MOLLI at 5% significance level (P=0.08). The mean and standard deviation of septal $T_1$ values between MOLLI and the proposed method across subjects is compared in Supporting Information Table S1.

Figure 9 shows representative $T_1$ and $B_1^+$ maps from Subject 2 obtained by proposed method. Notice the similarity in estimated $T_1$ values for each tissue type (e.g., myocardium, liver, and muscle) even with significant variations in estimated $B_1^+$ across different regions (Fig. 9a). The 3D $B_1^+$ maps of the heart (Fig. 9b) show larger $B_1^+$ values in lateral/anterior regions than septal/inferior regions, which is consistent with literature[39]. Group analysis of $B_1^+$ maps acquired from all subjects are shown in Fig. 10. $B_1^+$ variation ranged from 21-32% and 11-15% across the septal-lateral and inferior-anterior regions of the myocardium in the left ventricle, respectively. When such $B_1^+$ inhomogeneities were ignored in $T_1$ estimation, $T_1$ values in the septal and inferior regions were overestimated (Supporting Information Figure S4). This was consistent with simulation results in Figure 2.

## Discussion

In this work, a new free-breathing cardiac $T_1$ mapping method is proposed for robust $T_1$ mapping of the heart at 3T. The $T_1$ maps obtained using proposed method have strong correlation and good agreement compared to reference and comparison methods in both phantom experiments with various conditions (Fig.4 and Supporting Information Fig.S2) and in vivo experiments across all subjects (Figs.6, 8 and Supporting Information Fig.S3). Robust $T_1$ mapping was achieved despite significant $B_1^+$ variations at 3T. This is most noticeable by the fact that uniform $T_1$ distributions for each tissue type (e.g., myocardium, liver, and muscle) were achieved across the entire FOV (Fig.9a) and is further supported by the close agreement of $T_1$ maps between proposed method and MOLLI (Fig.8), despite the observed $B_1^+$ variation across septal-lateral and inferior-anterior regions of the myocardium in the left ventricle (Fig.10). Noticeable bias in $T_1$ estimation was otherwise observed when $B_1^+$ inhomogeneities were ignored in $T_1$ estimation (Figs.2,4,8,10, Supporting Information Figs.S1 and S4). These observations were consistent throughout simulation, phantom, and in vivo results. The estimated $B_1^+$ values of the myocardium show spatial variations which are consistent with those reported in literature[39], i.e., the $B_1^+$ values in the lateral/anterior regions were 10-30% larger than the septal/inferior regions. However, the $B_1^+$ distributions within the blood pool should be carefully interpreted since flow effects were not considered in joint $T_1$ and FA estimation. As a result, the estimated $T_1$ and FA values of blood may be biased.

This method has several novel features. First, the proposed method mitigates bias in $T_1$ estimation caused by errors in FA via joint estimation of $T_1$ and FA. We carried out systematic numerical simulation studies to optimize the ECG-gated IR sequence with SPGR readout in terms of acquisition protocols and nominal FAs with the goal of minimizing the standard deviation of the estimated $T_1$. Second, the proposed method utilizes special $(k,t)$-space sampling scheme and subspace-based image reconstruction to recover dynamic images from under-sampled data, *i.e.,* two 3D volumes for every cardiac cycle with data acquisition. This allows mitigating the effects of natural heart-rate variations on $T_1$ mapping by fitting the reconstructed dynamic signals to a signal dictionary generated with subject-specific timing of data acquisition recorded during imaging experiment. Third, the proposed method is robust to $B_0$ inhomogeneities since it utilizes adiabatic non-selective pulse for inversion and SPGR acquisitions. Altogether, the proposed method achieves free-breathing $T_1$ mapping in the presence of $B_1^+$ and $B_0$ inhomogeneity at 3T in a practical imaging time.

The proposed method may be potentially useful for quantification of post-contrast $T_1$ and ECV mapping where accurate and precise estimation of $T_1$ is important. Although results from simulation and phantom studies show that proposed method can estimate short $T_1$ values with accuracy and precision, further investigation is necessary to evaluate the performance of proposed method for post-contrast $T_1$ estimation. Since $T_1$ relaxation in the tissue changes over time in vivo after contrast agent is injected, the performance of proposed method needs to be carefully examined for these applications, including investigations in the context of subspace-based reconstruction. An interesting next step would be to investigate the feasibility as well as the performance of the proposed method for these applications in vivo.

In this work, a subspace-based image reconstruction method was used to recover dynamic images in clock time for ECG-gated acquisitions and respiratory motion was resolved by subsequently binning reconstructed images to different respiratory motion phases. A potentially interesting future work would be to investigate possibility of treating respiratory motion as an additional temporal dimension using low-rank tensor model(16) for ECG-gated acquisitions. In the ideal case with fixed heartrate, this would be feasible since $T_1$-weighted contrast changes can then be modeled by inversion recovery at a small number of inversion delay times. In reality, however, natural variations of heartbeat rate will require modeling the $T_1$-weighted contrast changes in clock time for ECG-gated acquisitions over the entire imaging experiment. This can impose technical challenges when attempting to represent the underlying dynamic images using a low-rank tensor model.

Results from in vivo study showed that $T_1$ values from proposed method were not statistically different from MOLLI at 5% significance level (P=0.08). This indicates that the proposed method has similar bias as MOLLI in in vivo study. Several factors contribute to the apparent underestimation of $T_1$ when proposed method is used for in vivo cardiac $T_1$ mapping. First, magnetization transfer effects can lead to underestimated $T_1$ values in the proposed method as in MOLLI since both methods use inversion-recovery based acquisition schemes for $T_1$ mapping. Second, the proposed method currently assumes perfect inversion pulse. Adding inversion efficiency to parametric fitting may potentially improve accuracy of $T_1$ estimation at the cost of computation time (i.e., a larger dictionary of bases needed for parametric fitting). Further investigation is necessary to study the potential source of bias in proposed method in in vivo settings.

The current work has several limitations that warrant further investigation. First, the proposed method involves ECG-gated acquisition, which is susceptible to ECG mis-triggering and may suffer from image blurring due to cardiac motion. The former could be addressed by adaptive heartbeat rate prediction as in double-gating technique(40). The latter can be mitigated by retrospectively discarding k-space data acquired outside the end-diastole window based on recorded ECG signals. Free-running (i.e., no cardiac or respiratory gating)-based continuous acquisition scheme may be preferable over ECG-gated acquisition schemes for maximizing data acquisition efficiency. However, in free-running continuous acquisition scheme, apparent inversion recovery rate is strongly coupled with FA. Therefore, accurate estimation of FA is expected to be critical for accurate $T_1$ estimation with free-running acquisition in the presence of $B_1^+$ inhomogeneity at 3T. This research direction is currently under investigation. Second, FA estimation by proposed method was validated using a phantom experiment with varying $B_1^+$ field strength via changing transmitter voltage (Supporting Information Fig.S2). Validation with a reference cardiac FA or $B_1^+$ mapping method, e.g., actual flip-angle imaging method(41), in human subject studies is necessary to further verify the performance of FA mapping results from proposed method. Third, performance of proposed method was validated in vivo with a small number of healthy subjects (n=6). Studies with a larger number of healthy subjects and patients are necessary to assess the accuracy and reproducibility of proposed method and to evaluate its value in clinical applications. Future work can also include investigation with different sparsity constraints (e.g., finite difference or wavelet transform), investigation with different sampling schemes, and multi-site/multi-vendor validations of proposed method and findings from this work.

## Conclusions

A new free-breathing cardiac $T_1$ mapping method was proposed and optimized for fast 3D $T_1$ mapping of the whole-heart at 3T with transmit B1 correction in practical imaging time.

## Acknowledgements

This work was supported in part by the National Institutes of Health (P41EB022544, R01CA165221, R01HL137230, R01HL118261, T32EB013180, and K01EB030045).

# Figure Captions

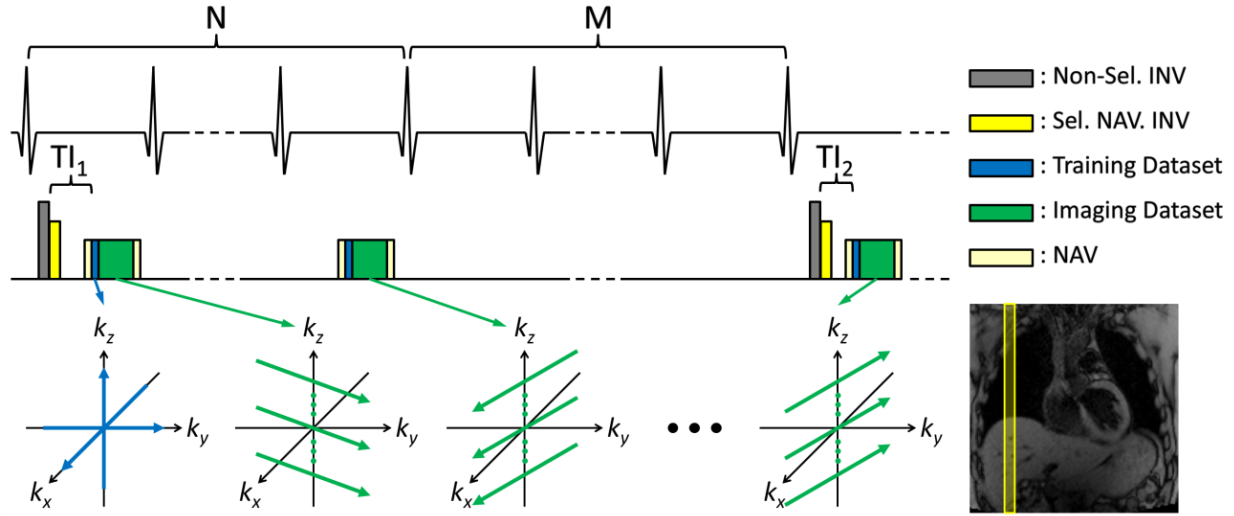

**Figure 1.** Schematic diagram of the proposed data acquisition scheme. N-(M)-N-(M) protocol is shown with non-selective inversion pulse applied every N+M heartbeats (where N denotes the number of cardiac cycles for acquisition and M denotes the number of cardiac cycles for signal recovery) with two different inversion times (TI). Data acquisition consists of "training" dataset acquiring a limited number of k-space lines with high sampling rate and "imaging" dataset sparsely-sampling all other k-space locations for subspace-based image reconstruction. To track respiratory motion, 1D respiratory navigator signals were acquired in the sagittal plane at the dome of the right hemi-diaphragm after a spatially-selective inversion pulse was applied in the same sagittal plane to invert the magnetization signals back to the equilibrium state.

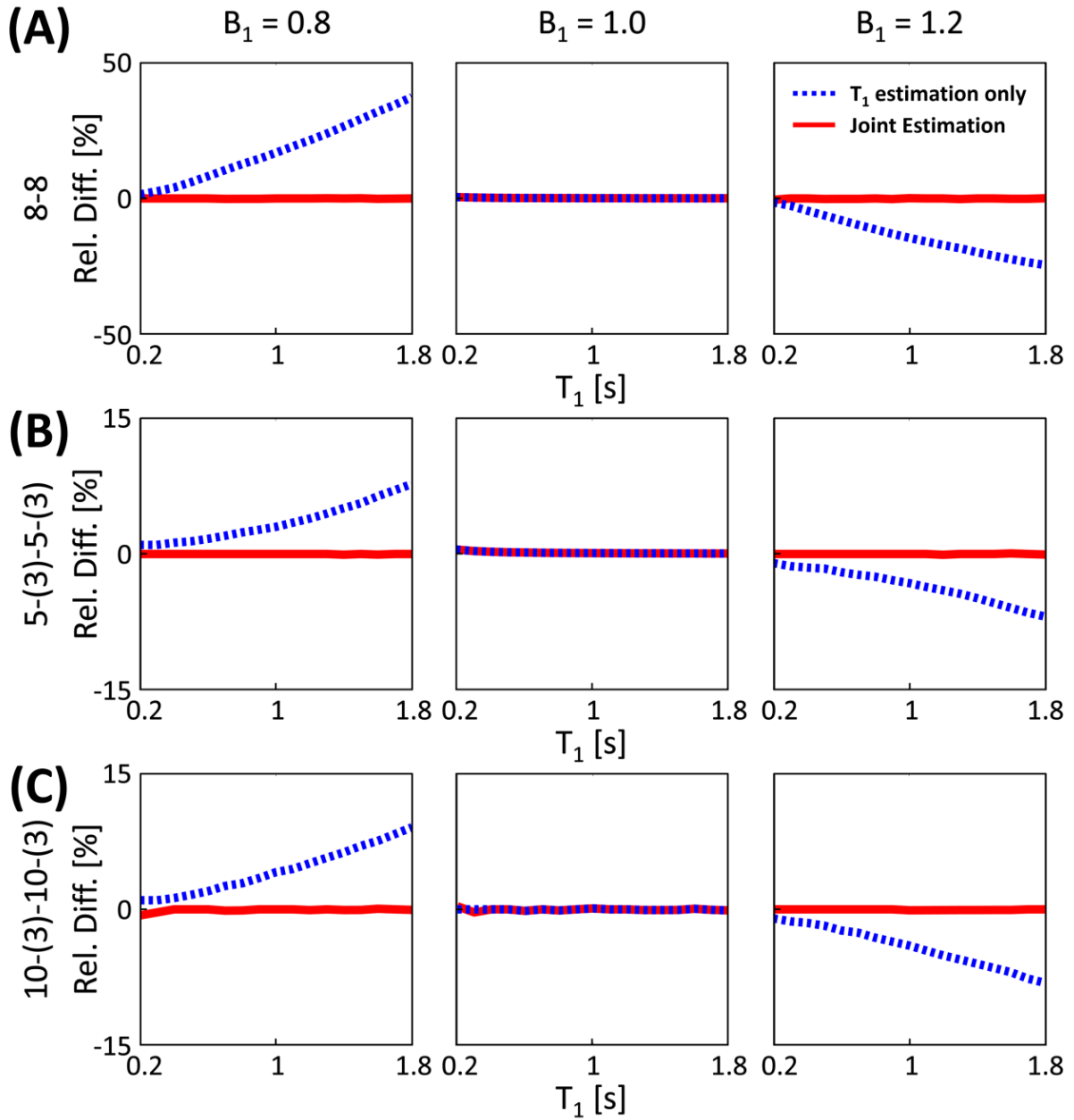

**Figure 2.** Simulation results showing the effect of $B_1^+$ inhomogeneity on the accuracy of $T_1$ estimation. Relative difference plots from the 8-8 (A), 5-(3)-5-(3) (B), and 10-(3)-10-(3) (C) protocols are shown for typical $B_1^+$ inhomogeneities at 3T (i.e., $B_1^+$ = 0.8, 1, and 1.2). $T_1$ values obtained by joint $T_1$ and FA estimation (red solid line) and $T_1$ estimation only with the assumption of perfect $B_1^+$ (blue dashed line) are shown. Small fluctuations are presumed to be due to numerical errors.

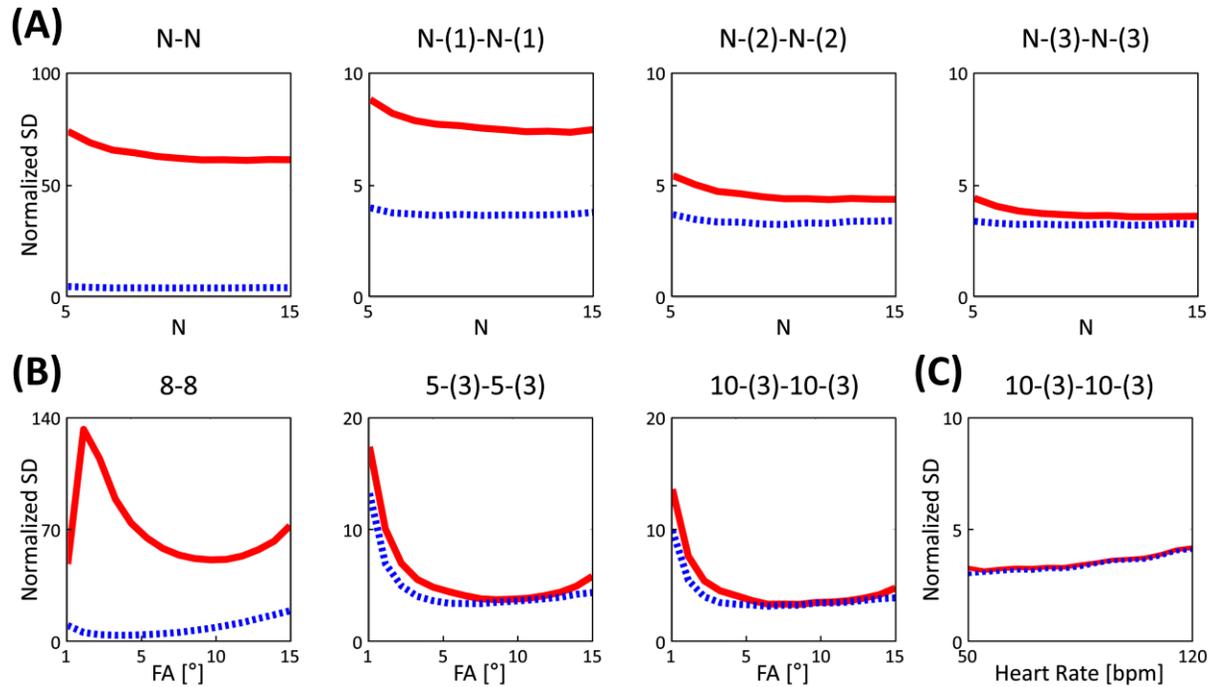

**Figure 3.** Simulation results showing the precision of $T_1$ estimation for different N-(M)-N-(M) protocols. **A**: Effect of N-(M)-N-(M) protocols on normalized standard deviation (nSD). **B**: Effect of flip angle (FA) on nSD for the 8-8, 5-(3)-5-(3), and 10-(3)-10-(3) protocols. **C**: Effect of heart rate variation on nSD for the 10-(3)-10-(3) protocol. $T_1$ values obtained by joint $T_1$ and FA estimation (red solid line) and $T_1$ estimation only with the assumption of perfect $B_1^+$ (blue dashed line) are shown.

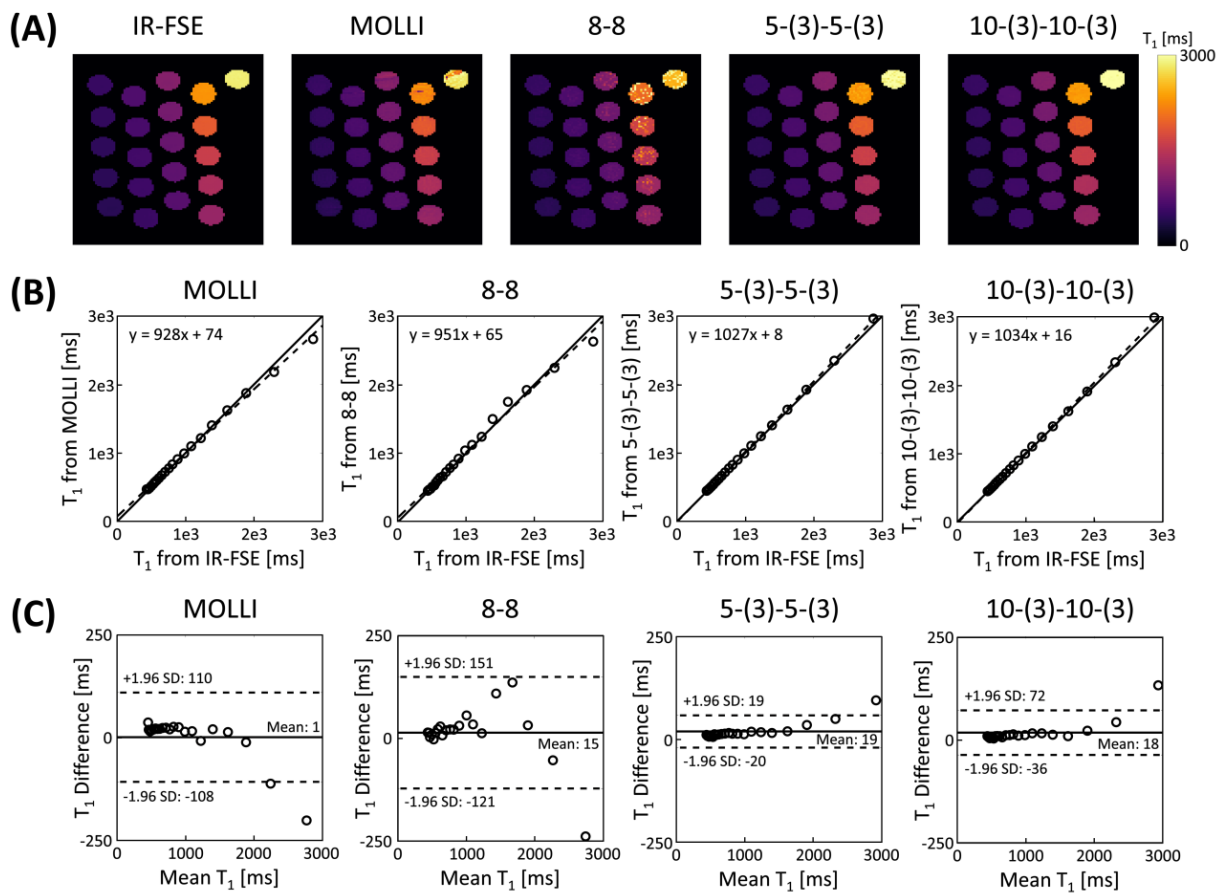

**Figure 4.** Phantom results of ECG-gated IR schemes with SPGR readout and joint $T_1$ and FA estimation. **A**: Estimated $T_1$ maps from different methods. **B**: Scatter plots showing comparison of estimated $T_1$ from different methods with those from IR-FSE. Solid line represents line of identity and dashed line represents line of regression. **C**: Bland-Altman plots showing comparison of estimated $T_1$ from different methods with those from IR-FSE. Solid line represents the mean difference and dashed line represents the 95% confidence interval for limits of agreement.

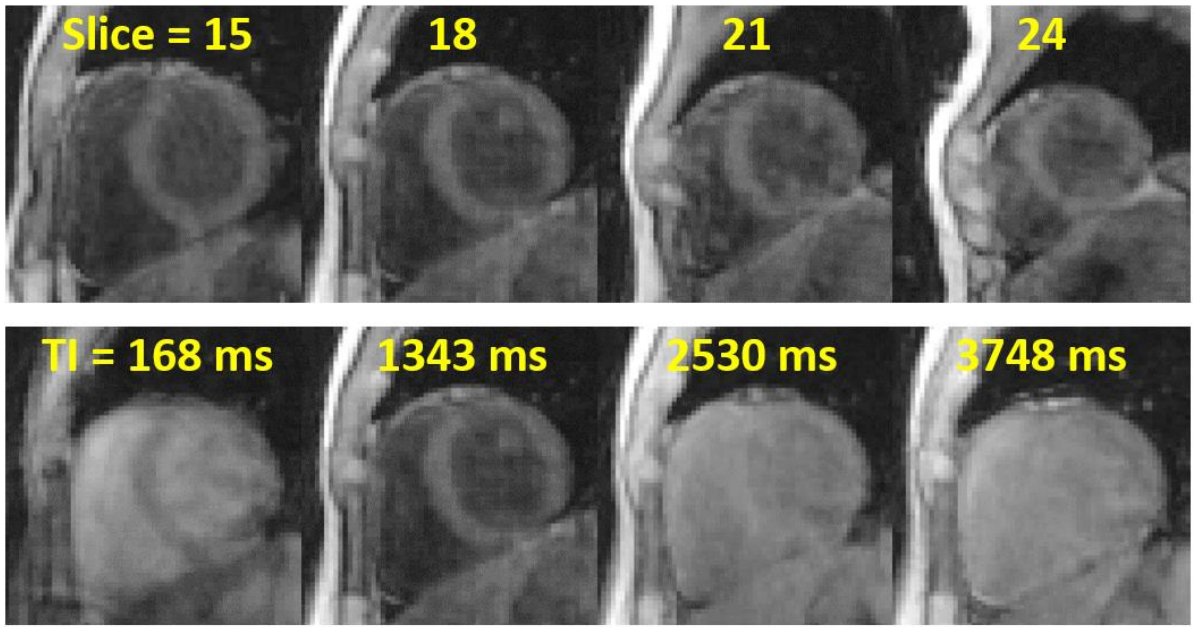

**Figure 5.** In vivo study results of reconstructed images from the proposed method. Representative reconstructed images from Subject 1 are shown at various slice positions for a fixed TI of 1343 ms (top row) and various TI times for slice position index of 18 (bottom row). Note that the images were selected from reconstructed images in clock time and may be at different respiratory motion phase.

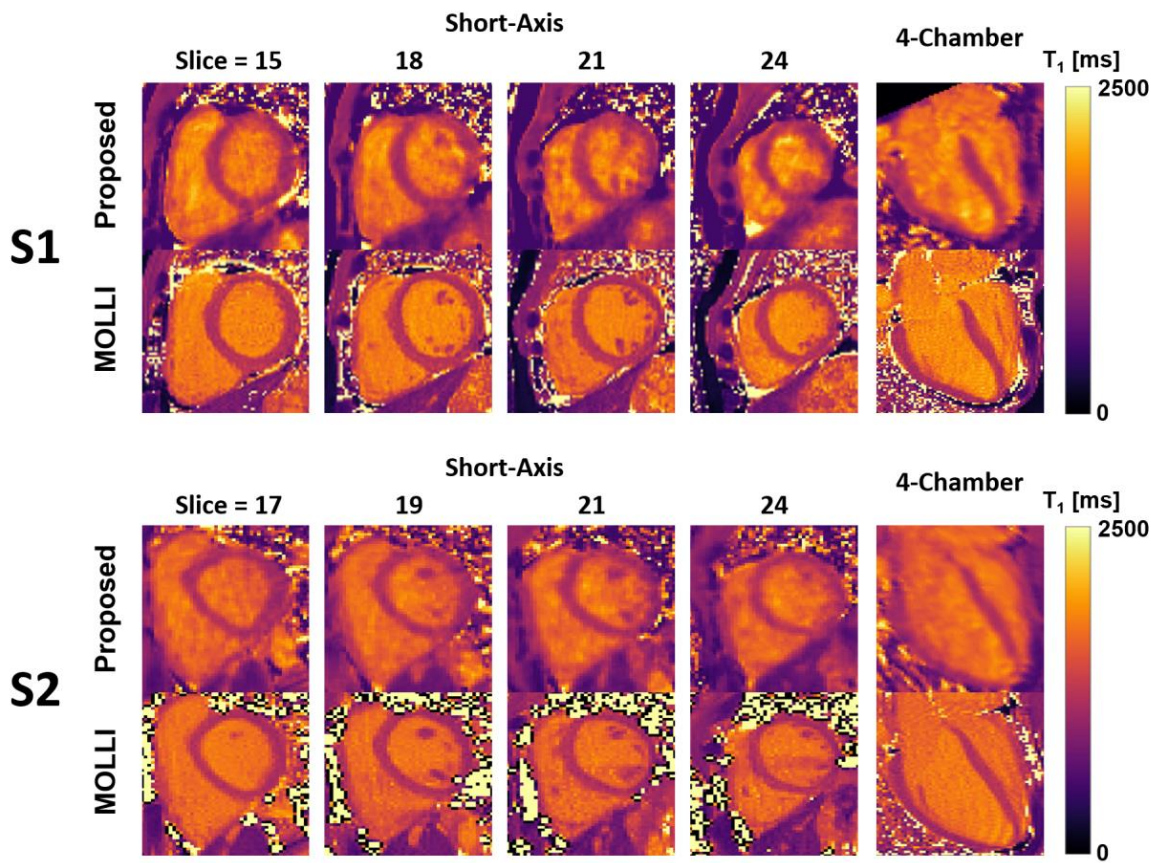

**Figure 6.** T$_1$ maps from Subject 1 and 2 obtained by the proposed method. Representative T$_1$ maps are shown for various slice positions in the short-axis view and 4-chamber view. T$_1$ maps from MOLLI at the same slice position and orientation are shown at the bottom of each subfigure for comparison. Note that for the 4-chamber view, the T$_1$ map from MOLLI were acquired with in-plane resolution of 1.5 mm, whereas the T$_1$ map from the proposed method was generated by re-slicing the 3D T$_1$ map acquired with through-plane resolution of 4.5 mm.

**Base** ⟶ **Apex**  T$_1$ [ms]

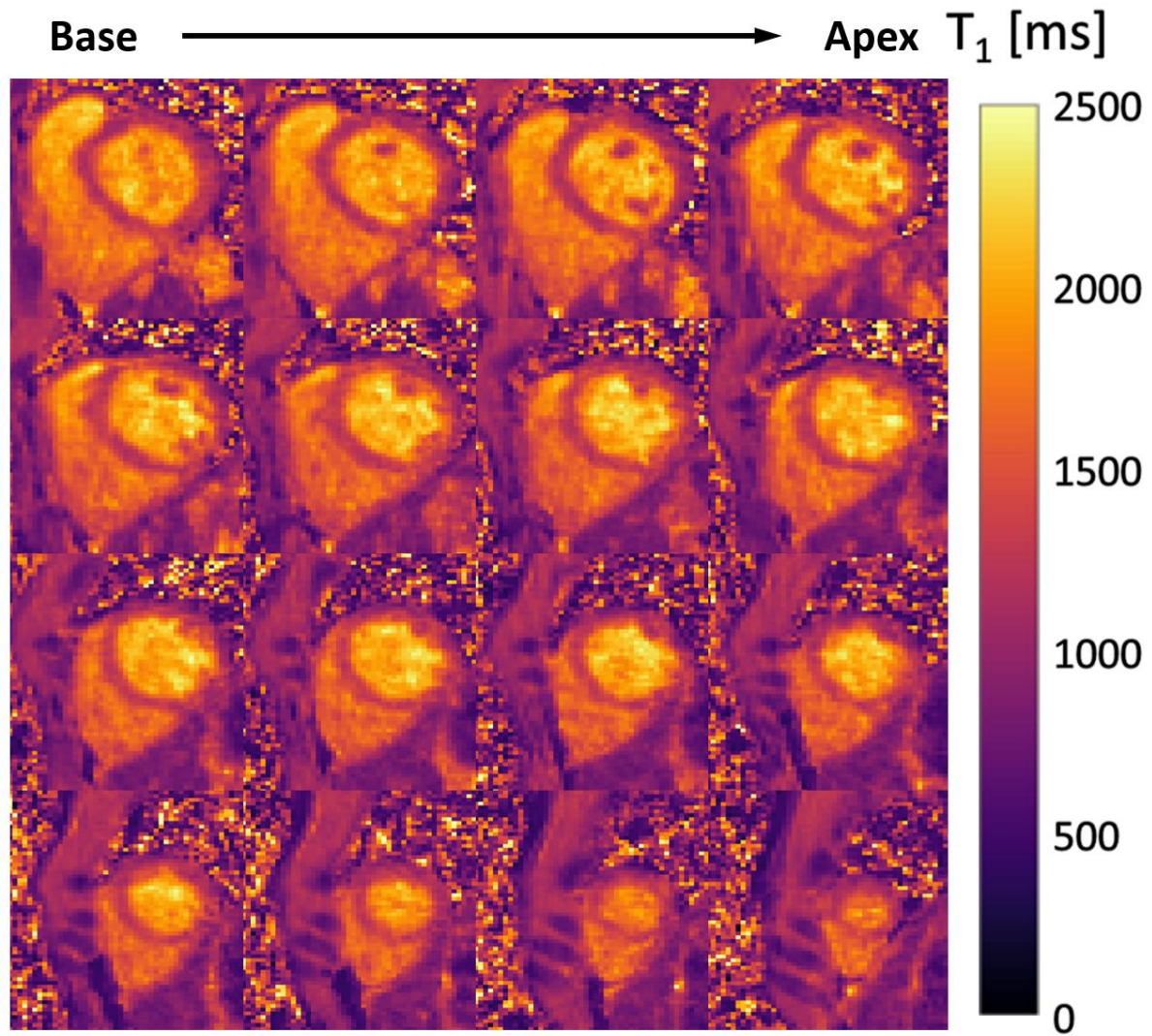

**Figure 7.** Representative 3D T1 map in the short-axis view covering the whole left ventricle from base to apex.

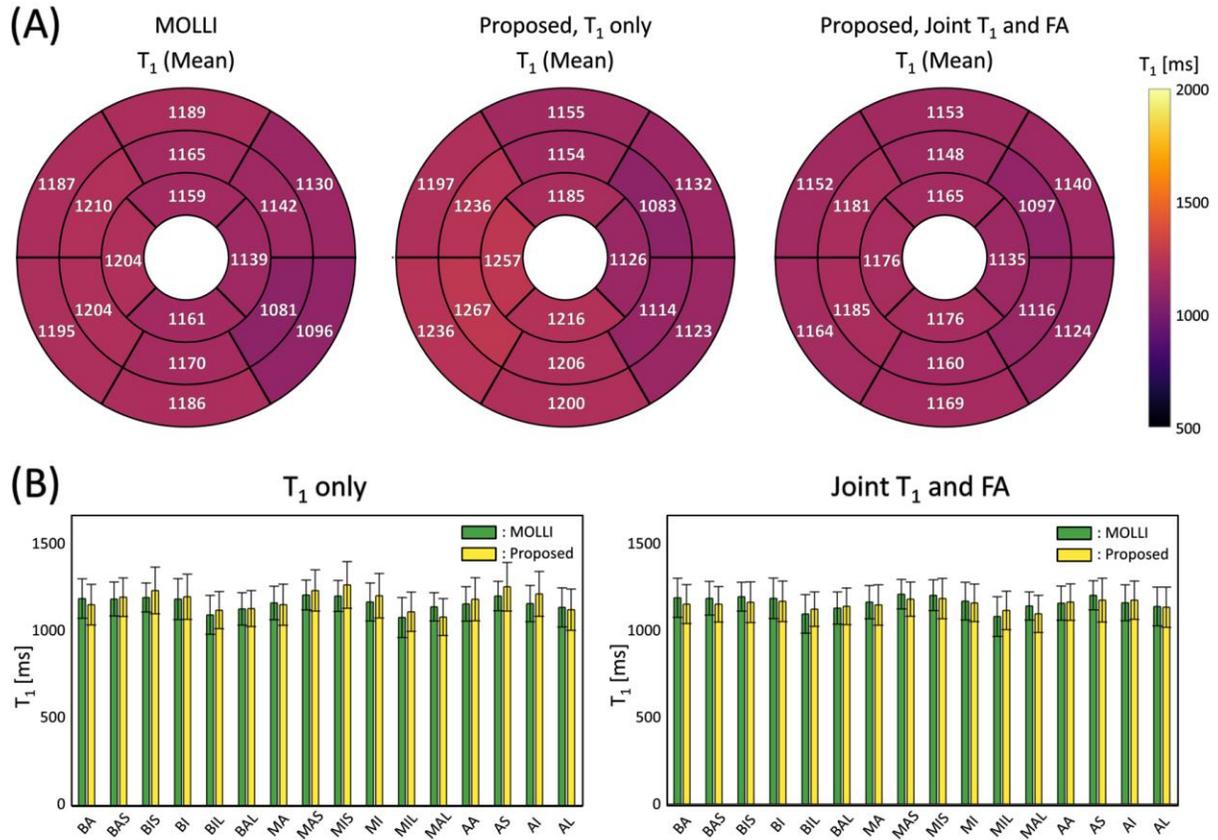

**Figure 8.** In vivo study results showing quantitative comparison of estimated $T_1$ from MOLLI and those from the proposed method. **A**: 16-segment AHA bull's eye plots of mean $T_1$ from MOLLI, proposed method with $T_1$ estimation only assuming perfect $B_1^+$ (i.e., $B_1^+ = 1$), and proposed method with joint $T_1$ and FA estimation. **B**: Bar plots showing mean and standard deviation of estimated $T_1$ from MOLLI and proposed method. Cases of proposed method with $T_1$ estimation only assuming perfect $B_1^+$ (i.e., $B_1^+ = 1$) and joint $T_1$ and FA estimation are shown. BA, BAS, BIS, BI, BIL, BAL, MA, MAS, MIS, MA, MIL, MAL, AA, AS, AI, AL each denotes basal anterior, basal anteroseptal, basal inferoseptal, basal inferior, basal inferolateral, basal anterolateral, mid-cavity anterior, mid-cavity anteroseptal, mid-cavity inferoseptal, mid-cavity inferior, mid-cavity inferolateral, mid-cavity anterolateral, apical anterior, apical septal, apical inferior, and apical lateral regions of the myocardium in the left ventricle.

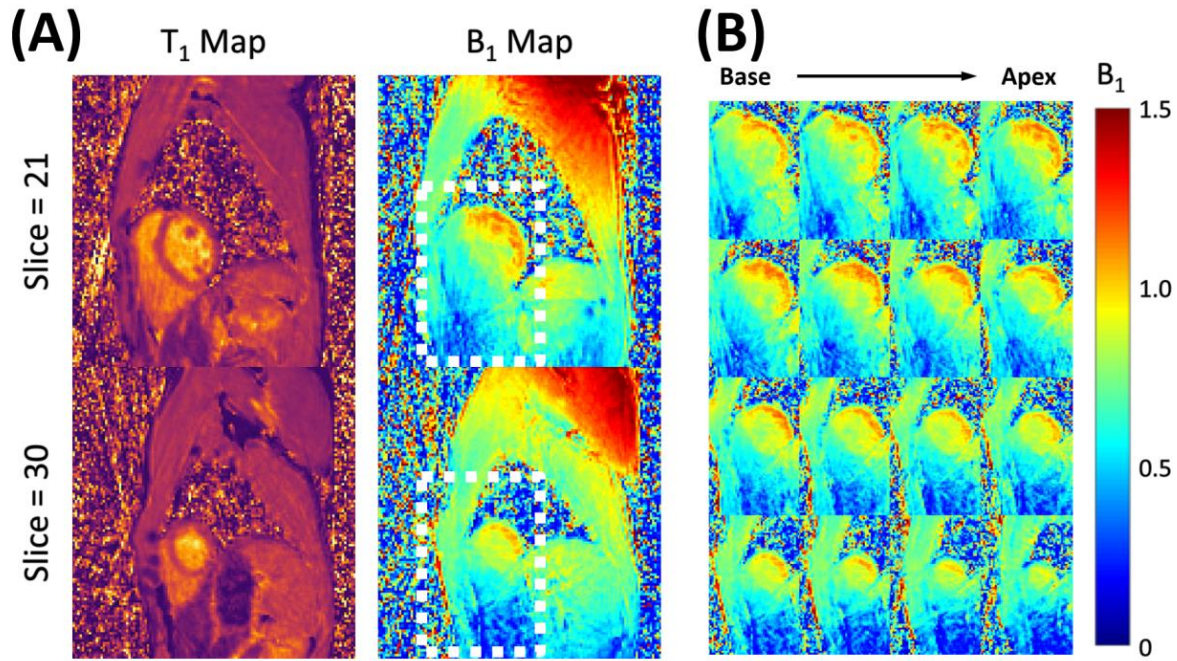

**Figure 9.** Representative 3D $B_1^+$ map (defined as the ratio between the measured and nominal FAs) obtained by the proposed method. **A**: Short-axis view $T_1$ and $B_1^+$ maps at two slice positions. Notice the smooth variation of $B_1^+$ across regions. Also notice the similarity in estimated $T_1$ for each tissue type even with variation in the estimated $B_1^+$ across different regions. **B**: Short-axis view $B_1^+$ maps of the heart at various slice positions from base to apex.

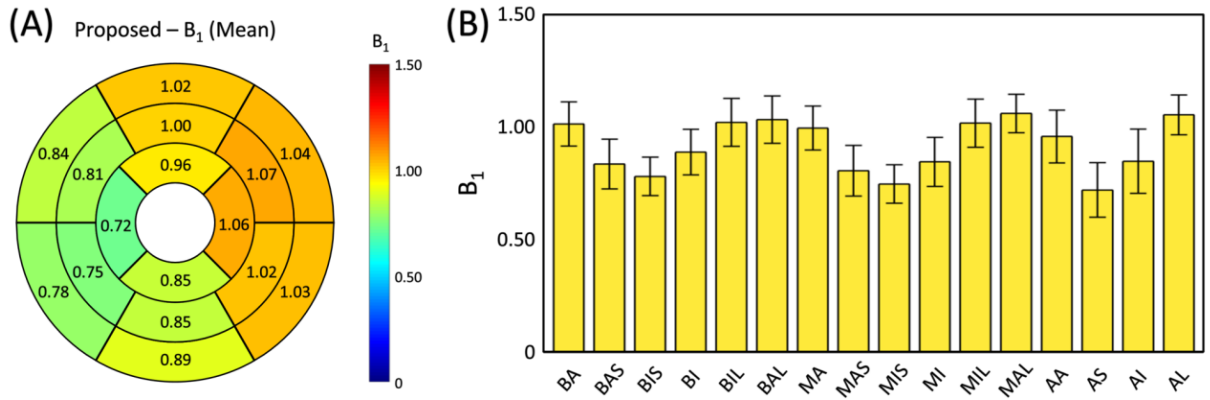

**Figure 10.** In vivo study results of estimated $B_1^+$ (defined as the ratio between the measured and nominal FAs) from all the subjects. **A**: 16-segment AHA bull's eye plot of normalized mean $B_1^+$ (i.e., $B_1^+$ normalized by the mean $B_1^+$ estimated from the mid-cavity anterior region of the myocardium in the left ventricle for each subject) estimated from the proposed method. Notice the difference in estimated $B_1^+$ between the septal-lateral and inferior-anterior regions of the myocardium. **B**: Bar plot showing mean and standard deviation of normalized $B_1^+$ estimated from the proposed method.

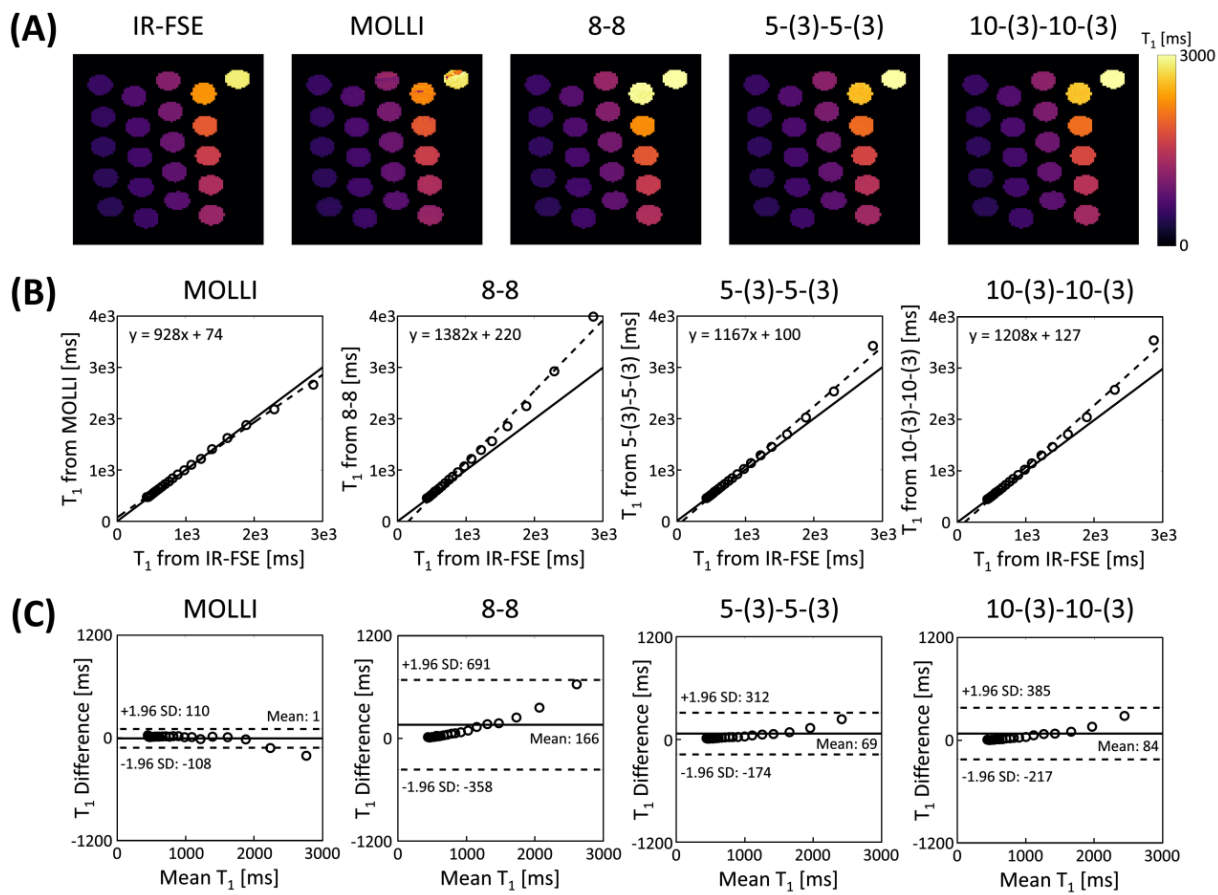

**Figure S1.** Phantom results of ECG-gated IR schemes with SPGR readout and $T_1$ estimation only assuming perfect $B_1^+$ (i.e., $B_1^+ = 1$). **A**: Estimated $T_1$ maps from different methods. **B**: Scatter plots showing comparison of estimated $T_1$ from different methods with those from IR-FSE. Solid line represents line of identity and dashed line represents line of regression. **C**: Bland-Altman plots showing comparison of estimated $T_1$ from different methods with those from IR-FSE. Solid line represents the mean difference and dashed line represents the 95% confidence interval for limits of agreement.

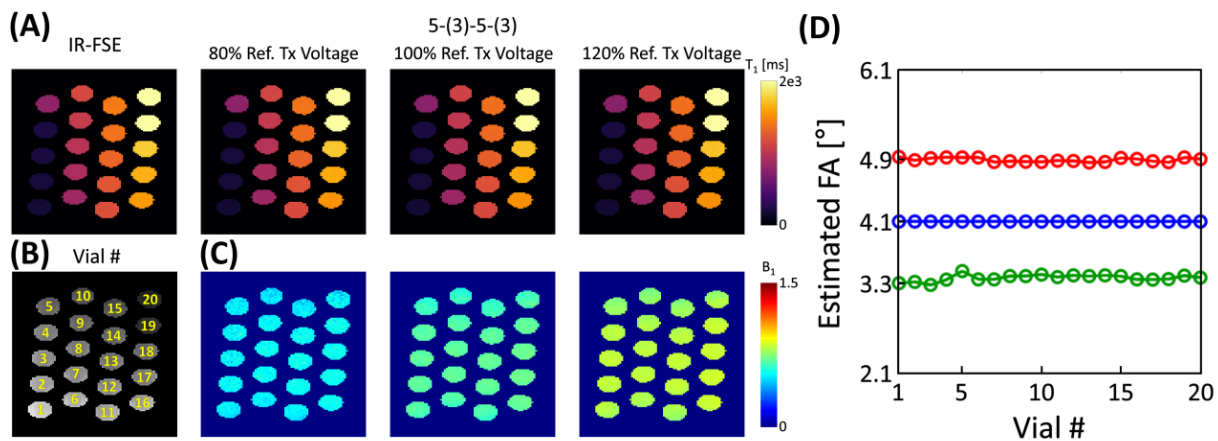

**Figure S2.** Phantom study results with variation in $B_1^+$ field strength via control of transmitter voltage. **A**: Estimated $T_1$ maps from IR-FSE and 5-(3)-5-(3) protocol with 80, 100, and 120% of reference transmitter voltage. **B**: Vial number positions. **C**: Estimated $B_1^+$ maps (defined as the ratio between the measured and nominal FAs) from 5-(3)-5-(3) protocol with 80, 100, and 120% of reference transmitter voltage. **D**: Estimated average FA within each vial from 5-(3)-5-(3) protocol with 80 (green line and circle), 100 (blue line and circle) and 120% (red line and circle) of reference transmitter voltage. Note that the ratio between the different $B_1^+$ field strength and the estimated average FAs from each vial are in good agreement.

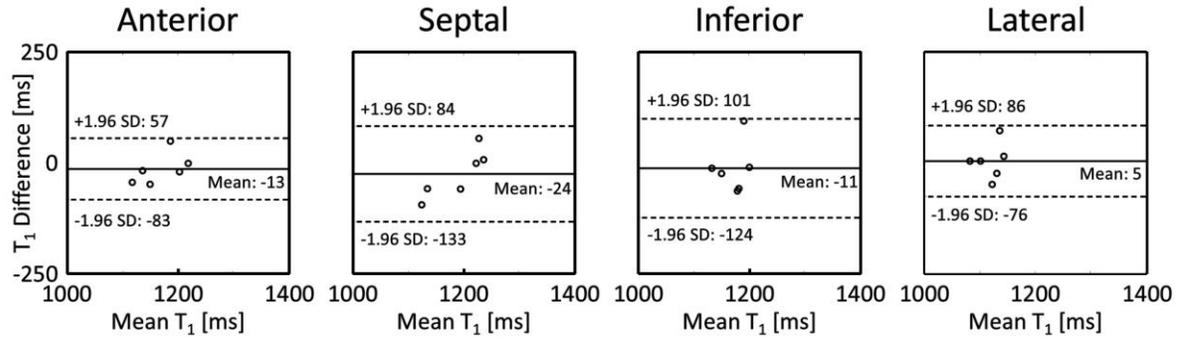

**Figure S3.** In vivo study results of Bland-Altman plots comparing estimated $T_1$ from the proposed method with those from MOLLI. Results from the anterior, septal, inferior, and lateral regions of the myocardium in the left ventricle are shown. Solid line represents the mean difference and dashed line represents the 95% confidence interval for limits of agreement.

**Table S1.** Comparison of mean and standard deviation of septal $T_1$ values between MOLLI and proposed method across subjects

|  | MOLLI | Proposed |
|---|---|---|
| Subject 1 | 1223.6 ± 72.8 | 1165.2 ± 73.9 |
| Subject 2 | 1222.8 ± 74.8 | 1222.8 ± 106.2 |
| Subject 3 | 1199.6 ± 65.7 | 1256.0 ± 83.4 |
| Subject 4 | 1171.1 ± 94.3 | 1077.0 ± 95.5 |
| Subject 5 | 1163.2 ± 101.1 | 1105.6 ± 84.9 |
| Subject 6 | 1232.2 ± 87.6 | 1240.4 ± 87.8 |